\documentclass[11pt,preprint]{aastex}

\def\spose#1{\hbox to 0pt{#1\hss}}
\def\simlt{\mathrel{\spose{\lower 3pt\hbox{$\mathchar"218$}}
     \raise 2.0pt\hbox{$\mathchar"13C$}}}
\def\simgt{\mathrel{\spose{\lower 3pt\hbox{$\mathchar"218$}}
     \raise 2.0pt\hbox{$\mathchar"13E$}}}
\usepackage{epsf}

\shorttitle{Dynamics \& Stellar Content of Giant M31 Stream}
\shortauthors{Font et~al.}

\begin{document}

\title{Dynamics and Stellar Content of the Giant Southern Stream in M31.
II.~Interpretation}

\author{
Andreea~S.~Font\altaffilmark{1},
Kathryn~V.~Johnston\altaffilmark{1},
Puragra~Guhathakurta\altaffilmark{2},
Steven~R.~Majewski\altaffilmark{3}, and
R.~Michael~Rich\altaffilmark{4}
}

\email{
afont@astro.wesleyan.edu,
kvj@astro.wesleyan.edu,
raja@ucolick.org,
srm4n@virginia.edu,
rmr@astro.ucla.edu
}

\altaffiltext{1}{Van Vleck Observatory, Wesleyan Univ., Middletown, CT 06459}
\altaffiltext{2}{UCO/Lick Observatory, Dept.\ of Astronomy \& Astrophysics,
Univ.\ of California, Santa Cruz, CA 95064}
\altaffiltext{3}{Dept.\ of Astronomy, Univ.\ of Virginia, Charlottesville,
VA 22903}
\altaffiltext{4}{Dept.\ of Physics \& Astronomy, Univ. of California, Los
Angeles, CA 90095}

\begin{abstract}
We examine the nature of the progenitor of the giant stellar stream in M31
using as constraints new radial velocity measurements of stream red giant
stars \citep[presented in the companion paper by][]{guh05} along with other
M31 data sets available in the literature.  We find the observations are best
fit by orbits that are highly eccentric and close to edge-on, with apo- to
peri-center ratios of order~25--30, and with apocenters at or only slightly
beyond the southern edge of the current data. Among these orbits, we are
able to find a few that plausibly connect the stream with the northern spur or with
the low-surface-brightness feature of similar high metallicity as the stream
\citep[originally reported by][]{fer02} to the east of M31's center.
In the latter case, if the connection is real, then the eastern debris
should lie well in front of M31 near the apocenter of the orbit.  Both the width of the 
debris and velocity dispersion measurements imply a rough lower limit on the mass of
the progenitor of $10^8M_\odot$. We use this limit and our orbits to discuss which of M31's
satellites could be plausibly associated with the stream.  In addition, we
predict that the width of the stream should increase beyond the southern edge
of the current data around the apocenter of the orbit and that the line-of-sight
velocity dispersion should exhibit significant variations along the stream.

\end{abstract}


\section{Introduction}\label{sec:intro}

The recently discovered giant stellar stream to the south of the Andromeda
spiral galaxy (henceforth referred to as the `giant southern stream' in M31)
is thought to be
debris from the (ongoing or past) disruption of a satellite dwarf galaxy
\citep{iba01a,fer02}.  This finding has sparked a series of followup
observations \citep {mcc03,iba04}, including those presented in the companion
paper by \citet[][hereafter Paper~I]{guh05}, as well as speculations about
possible associated objects \citep{mer03,mer05,hur04}.  Such extended debris is 
interesting because the dynamics are relatively simple to model
\citep{tre93,joh98,hel99}: The stars in debris streams are dissipationless,
so the essential ingredient of these models is simply phase-mixing along a
single orbit.  As a consequence, streams offer a potential goldmine of
information about their origins, with constraints on the orbit, mass, and
time since disruption of the progenitor object buried in the morphology and
kinematics of the debris \citep[see][for a general discussion of interpreting
streams around external galaxies]{joh01}.

The best studied example of satellite disruption is the Sagittarius dwarf
galaxy \citep*[discovered by][]{iba94}, a satellite of our own Milky Way
galaxy \citep[see][for a review of observational work]{maj03}.  Models for
the Sagittarius dwarf's debris have not only told us about its own history
\citep*[e.g.,][]{joh95,vel95,iba98,gom99,law05}, but also offered insights into the
shape of the Milky Way's potential \citep*{iba01b,joh05}.  Information
extracted from debris around other galaxies is in general much more limited
because the data sets are usually restricted to surface photometry, with no
practical way to measure distance or velocity variations.  The M31 stream is
the first example of debris around another galaxy that can be studied in more
detail because it is close enough that the individual giant stars can be
resolved, distances estimated from the tip of the red giant branch
\citep{mcc03}, and velocities obtained from spectra
\citep[Paper~I;][]{iba04}.  Such studies have already led to specific
estimates of the orbit of the progenitor, and limits on M31's mass
\citep{iba04}.  A more detailed investigation of the nature of the progenitor
is now possible with recently acquired data on the width and the velocity
dispersion of the stream.

In this paper, we revisit the constraints on the orbit of the progenitor
(\S\,\ref{sec:orb}), estimate what limits can be placed on its mass
(\S\,\ref{sec:width}), and which M31 objects (and other 
low-surface-brightness features) could be plausibly associated with the 
stream, given these estimates (\S\,\ref{sec:disc}). We summarize our 
conclusions in \S\,\ref{sec:summary}.

\section{Constraints on the Orbit}\label{sec:orb}

\subsection{Observational Constraints}\label{sec:obs_constr}

\begin{table}[ht!]
\begin{center}
\caption{Positional, line-of-sight distance, and radial velocity data (with
respect to M31) for fields along the giant southern stream and satellite
galaxies.  A colon indicates an uncertain measurement, and an ellipsis
indicates missing data.  The positions and line-of-sight distances for
fields~`1'--`13' are from \citet{mcc03} and the radial velocities for
fields~`1', `2', `6', and `8' are from \cite{iba04} and \citet{lew05}.  The
radial velocity for the field~`a3' is from our data (see Paper~I); for the
distance to this field, an intermediate value between fields~`4' and `5' is
adopted (see text for details).  The data for the satellites M32 and NGC~205
are from \citet{mat98}; the data for And~VIII are given by \citet{mor03}.
Here $(\xi, \eta)$ are the central coordinates of And~VIII, a feature 
which is found to extend approximately 10 kpc parallel to the semi-major axis 
of the M31 disk and about 2 kpc along the semi-minor axis, respectively. A
distance of 780~kpc is adopted for M31 as in \citet{mcc03} for consistency
with the distance determinations of their stream fields.}
\begin{tabular}{rrrll}
 & & & & \\
\hline\\
Field/Name  & $\xi~~~\,$ & $\eta~~~\,$ & $~~~d$ & $v_{\rm rad}$ (with respect to M31) \\
            & (deg)~     & (deg)~      & (kpc)  & ~~~~~(km~s$^{-1})$  \\
\hline
\hline\\
      a3~~~~& $+1.077$   & $-2.021$    & ~~850: & ~~~~~$-158$         \\
\hline\\
       1~~~~& $+2.015$   & $-3.965$    & ~~886  & ~~~~~$\>\>$~~~~0:   \\
       2~~~~& $+1.745$   & $-3.525$    & ~~877  & ~~~~~$\>\>-50$:     \\
       3~~~~& $+1.483$   & $-3.087$    & ~~860  & ~~~~~~~~...         \\
       4~~~~& $+1.226$   & $-2.653$    & ~~855  & ~~~~~~~~...         \\
       5~~~~& $+0.969$   & $-2.264$    & ~~840  & ~~~~~~~~...         \\
       6~~~~& $+0.717$   & $-1.768$    & ~~836  & ~~~~~$-180$:        \\
       7~~~~& $+0.467$   & $-1.327$    & ~~829  & ~~~~~~~~...         \\
       8~~~~& $+0.219$   & $-0.886$    & ~~780  & ~~~~~$-300$:        \\
      12~~~~& $-0.731$   & $+0.891$    & ~~739  & ~~~~~~~~...         \\
      13~~~~& $-0.963$   & $+1.342$    & ~~758  & ~~~~~~~~...         \\
\hline\\
     M31~~~~& 0.0~~~     & 0.0~~~      & ~~780  & ~~~~~~~~$\,\,\>\>$0 \\
     M32~~~~& 0.0~~~     & $-0.4~~~$   & ~~780  & ~~~~~$+100$         \\
 NGC~205~~~~& $-0.5~~~$  & $+0.4~~~$   & ~~830  & ~~~~~~$\,+55$       \\
And~VIII~~~~& 0.1~~~     & $-0.5~~~$   & ~~~... & ~~~~~$-204$     \\
\hline
\end{tabular}
\label{tab:ic_data}
\end{center}
\end{table}

The available spatial and velocity information on the giant southern stream
and satellites of M31 are discussed in this section and summarized in
Table~\ref{tab:ic_data}.  In the following section, these data are used to
set up the initial conditions and to serve as additional constraints for our
orbit integration.

Our own observations (Paper~I) provide two important constraints on the orbit
of the giant southern stream:
\begin{itemize}
\item[(1)]{The mean radial velocity in the field~`a3' is $-458$~km~s$^{-1}$
relative to the Sun which translates to $v_{\rm rad}-158$~km~s$^{-1}$ with
respect to M31.}
\item[(2)]{The position-velocity data provide a measure of the velocity
gradients both along the stream and perpendicular to it:
$dv/dr_\parallel\sim-0.5$~km~s$^{-1}$~arcmin$^{-1}$ and
$dv/dr_\perp\sim+0.6$~km~s$^{-1}$~arcmin$^{-1}$, respectively.  There is
considerable uncertainty, however, in the determination of these slopes due
to possible confusion between stream stars and those in the smooth M31 halo
as well as small number statistics.}
\end{itemize}

Our data are complemented with information from a few other studies:
\begin{itemize}
\item[(1)]{\citet{mcc03} have estimated line-of-sight distances as a function
of sky position for several fields along the giant southern stream.  These
imply that the stream sweeps from over 100~kpc behind M31 at the point
furthest from the disk in the south ('field 1') to 30~kpc in front of the
disk in the north (field~`13'---see Table \ref{tab:ic_data} for a summary).}
\item[(2)]{\citet{iba04} and \citet{lew05} find that the southernmost tip of
the stream is nearly at rest with respect to M31 (i.e.,~moving at the
systemic velocity of M31), whereas the stream in the vicinity of the disk
reaches a radial velocity of about $-300$~km~s$^{-1}$ with respect to M31.
This difference of about 165~km~s$^{-1}$ in radial velocity between
fields~`1' and `6', subtending about $3^\circ$ across the southern part of
the stream \citep[see Fig.~1 of][]{iba04}, implies a velocity gradient of
$dv/dr_\parallel\simeq-0.9$~km~s$^{-1}$arcmin$^{-1}$, in rough agreement with
the observed value within our field~`a3' (see Fig.~8 of Paper~I).}
\end{itemize}

\begin{figure}
\epsfbox{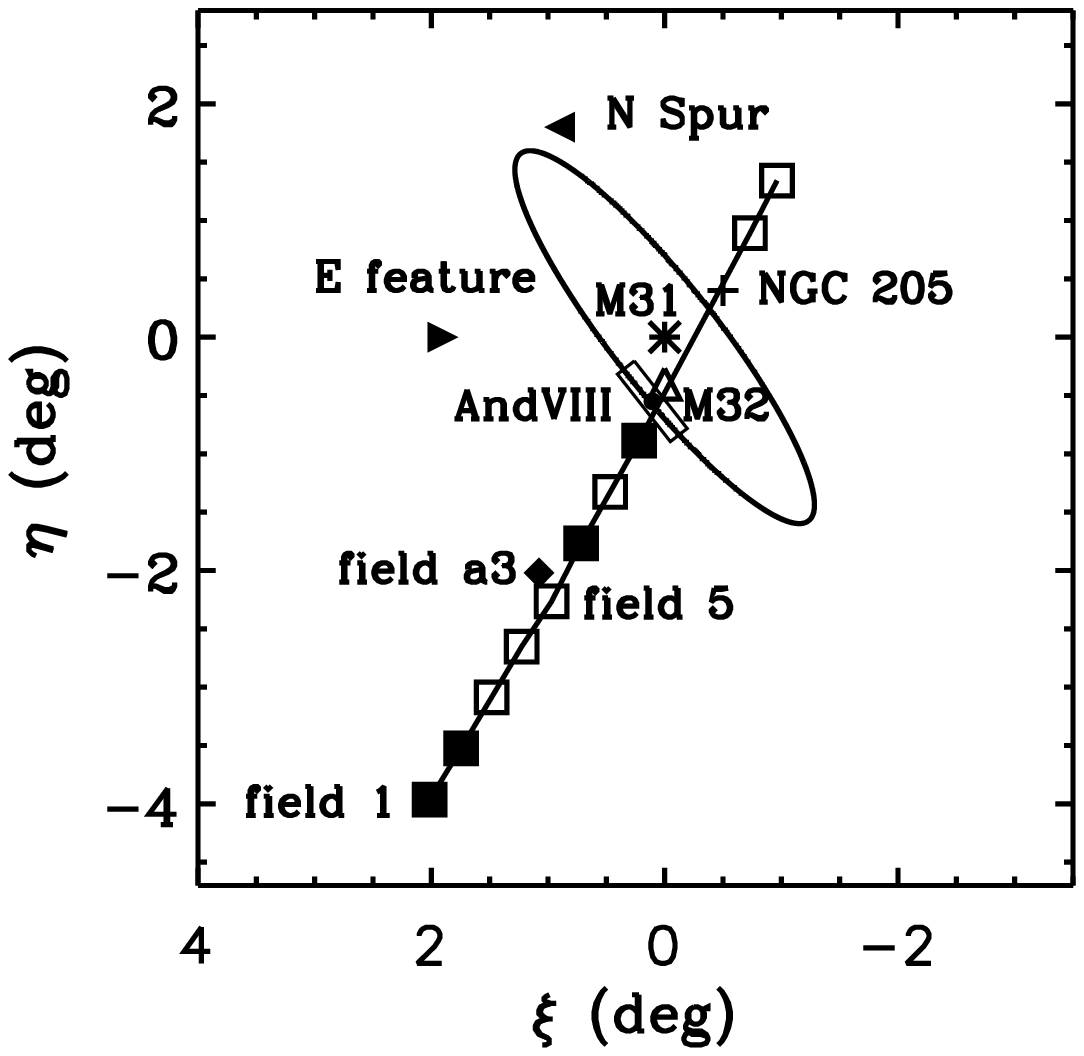}
\caption{\label{fig:obs_constr1}{The position of the stream fields, M31
satellite galaxies, and other stellar features expressed in standard
coordinates $\xi$ and $\eta$.  The filled squares and diamond denote the
fields for which radial velocity information is avalable (see
Table~\ref{tab:ic_data}).  The ellipse delineates the limit of the visible
disk of M31 with a semi-major axis length of $2^\circ$
\citep[see][]{fer02}.  The line connecting the
stream fields traces the extent of the giant southern stream, as detected so
far.  The eastern high-metallicity feature lies roughly at $\xi\sim2^\circ$
and $\eta\sim0^\circ$.  The narrow rectangular strip delineates And~VIII, a feature 
which is found to extend approximately 10~kpc parallel to the major axis
of M31 disk and about 2~kpc along the minor axis \citep{mor03}.
}}
\end{figure}

Figure~\ref{fig:obs_constr1} illustrates the positional data of the giant
southern stream fields and of M31's satellite galaxies.  The spatial and
velocity information together offer a general picture of the dynamics of the
stream. The southern part of the stream is located behind the disk (as seen
from our location) and is travelling generally towards M31 along almost a
straight line path, with an inclination of about $60^\circ$ with respect to
the line of sight \citep{mcc03}---this implies that the orbital plane must be
inclined by at least $i=30^\circ$ to the plane of the sky and cannot be face
on.

Indeed, the linearity of the stream in the sky and its proximity to the
center of M31 in field~`8' suggests that the inclination\footnote{Throughout
the paper, the inclination $i$ is used to denote the angle between the orbital plane and the
plane of the sky.  An inclination angle of $i=0^\circ$ corresponds to the case
in which the normal vector $\vec{n}$ to the orbit---defined by the direction of the total
angular momentum of the orbit---is oriented towards the observer.} of the orbital plane
is closer to $i=90^\circ$ (i.e.,~edge on) since any curvature of the orbit
would otherwise be apparent.  In addition, the strong velocity and distance
gradients along the stream and the large range in measured separations from
M31 along the stream indicate that the orbit is eccentric.  Lastly, the
linearity of the stream suggests that its orientation in space corresponds to
the direction of motion, with negligible motion perpendicular to it.  Hence,
the full space velocity relative to M31 can be estimated at each point along
the observed stream to be of order $v_{\rm rad}/\cos(60^\circ)=2v_{\rm rad}$.
Since the southernmost part of the stream has a radial velocity close to zero
with respect to M31, this would imply that this location is near or even
coincides with the apocenter of the orbit.

In addition to the above data, there are two~other ``features'' that stand
out in the star-count and metallicity maps:
\begin{itemize}
\item[(1)]{A high density stellar feature is observed near the northeastern end
of the disk major axis \citep{fer02}, and is known as the ``northern spur''.
Its metallicity is higher than that of the disk and the neighboring halo.
The origin of this feature is still unknown.  It has been hypothesized to be
either part of the giant southern stream or an extension of the disk
\citep{fer02,mer03}.  In the latter case, this would imply a very
significant warp of the disk.}
\item[(2)]{The high metallicity feature noted by \citet{fer02} immediately to
the south of the northeastern half of the disk.  Surprisingly, the
metallicity of this feature is comparable to that of the giant southern
stream.  It also appears to be higher than the overall metallicity of the
northern spur \citep[see Fig.~5 of][]{fer02}---although we caution that it is
possible that a similar high-metallicity component may be present in the
northern spur but may be hard to disentangle from the large number of typical
(i.e.,~lower metallicity) halo
stars in the region.  Since it lies more or less east of M31's center, we refer to
it as the ``eastern high-metallicity feature''.}
\end{itemize}
\noindent
The peculiarities of these two~features raise the question of whether they
may be related to the orbit of the giant southern stream.  Unfortunately, no
distance or velocity measurements are available yet for either of these
features and they are therefore not included as constraints in the orbit
integrations.  However, their possible connection with the stream, as
inferred from our orbit integrations, is discussed later in the paper
(\S\,\ref{sec:impl}).

\subsection{Test Particle Orbits}\label{sec:part_orb}

We now integrate test particle orbits in a static M31 potential in order to
find the general characteristics of those that could be consistent with the
data summarized in Table \ref{tab:ic_data}. The form of the potential contains
three components: a dark halo,

\begin{equation}
\Phi_{\rm halo}\,=\,v_{\rm halo}^2\log(r^{2} + d^{2})~~,
\label{eqn:pothalo}
\end{equation}

\noindent a \cite{miya75} disk,

\begin{equation}
\Phi_{\rm disk}\,= - \frac{G \, M_{\rm disk}}{\sqrt{R^{2} + (a + \sqrt{z^{2} + b^{2}})^{2}}}~~,
\label{eqn:potdisk}
\end{equation}

\noindent and a \cite{hernq90} bulge,

\begin{equation}
\Phi_{\rm bulge}\,= \frac{G \, M_{\rm bulge}}{r + c}~~.
\label{eqn:potbulge}
\end{equation}

\noindent
For the parameters in relations (\ref{eqn:pothalo})--(\ref{eqn:potbulge}),
we adopt the same values that \cite{bekki01} used to obtain a reasonable fit to
the M31 data: $d=12$~kpc, $v_{\rm halo}=131.5$~km~s$^{-1}$, $M_{\rm disk}=
1.3\times10^{11}\,M_{\odot}$, $a=6.5$ kpc, $b=0.26$ kpc, $M_{\rm bulge}
= 9.2\times10^{10}\,M_{\odot}$, and $c=0.7$ kpc.  With these parameters the
rotation speed reaches 260 km~s$^{-1}$ at a radial distance of 26 kpc, in
good agreement with the observations of M31 \citep[e.g.,][]{ken89} and with
the recent global mass constraint derived by \citet{iba04}.  Note that relation
(\ref{eqn:pothalo}) assumes that there is no flattening of the halo potential.
Present observational data are insufficient to probe the extent to which the
M31 halo may be flattened. However, given that the stream data are
confined to less than one orbital period, we expect the effects of weak or
moderate flattening to be minimal \citep[see also][]{mer03}.

Throughout the remainder of this paper, we denote with $(x,\,y)$ the
coordinates in the plane of the sky [aligned with angular coordinates
$(\xi,\,\eta)$] and $z$ along the line of sight.  The orbit integrations are
performed in a coordinate system having two axes aligned with M31's disk
(denoted  $x_{\rm M31}$ and $y_{\rm M31}$) and the third one, $z_{\rm M31}$,
perpendicular
to the disk.  We choose to start our orbit integrations from a location near
the center of the giant southern stream, where both spatial and velocity data are
available.  Thus, we choose field~`5' as starting point and assign to it the
radial velocity found in the neighboring field~`a3', $v_{\rm
rad}=v_{z}=158$~km~s$^{-1}$
(this choice is reasonable due both to the proximity of these two
fields, and because Fig.~8 of Paper~I shows that a strong gradient in the
direction perpendicular to the stream, $dv/dr_\perp$, can be ruled out).
Starting from field `5', we then integrate both backward and forward in time.

Given the lack of data for the other two components of the velocity, $v_{x}$
and $v_{y}$, we decided to construct a grid of orbits in order to constrain
this parameter space.  An initial inspection of the overall parameter
space shows that orbits which are good fits to the data have initial
conditions which cluster in the vicinity of the value $(v_{x},\,v_{y})=
(-80,\,132)$ km~s$^{-1}$.  Our final grid consists of $(21\times21)$ orbits,
all having a fixed initial radial velocity, $v_{z}=158$ km~s$^{-1}$ and
sampling the $(v_{x}$ and $v_{y})$ plane in steps of 5 km~s$^{-1}$ around
$(v_{x},\,v_{y})=(-80,\,132)$ km~s$^{-1}$.

\begin{figure}
\begin{center}
\epsfxsize=11cm\epsfysize=11cm\epsfbox{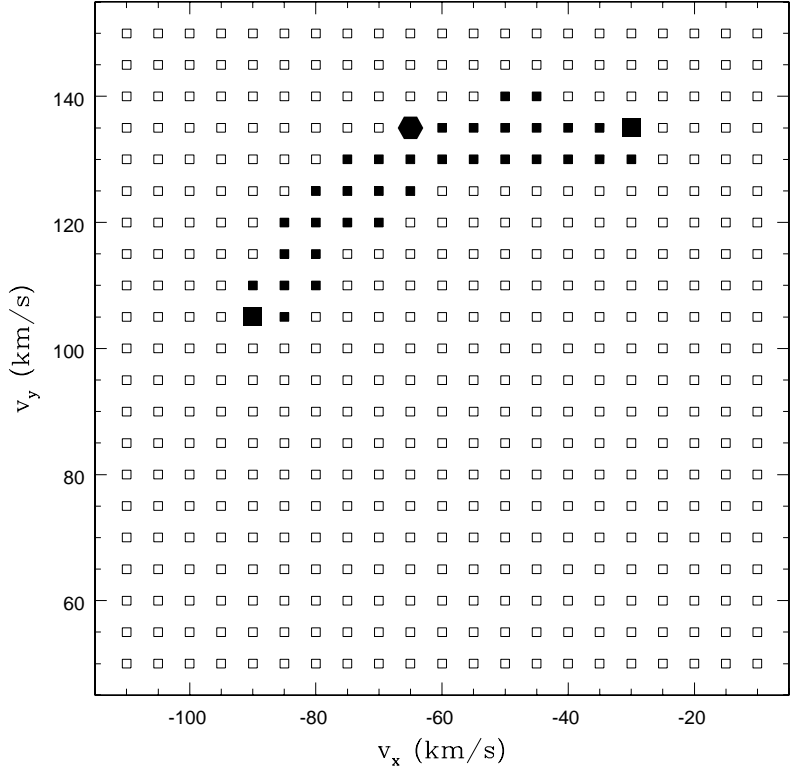}
\caption{\label{fig:grid}{Grid of $(21\times21)$ orbits with fixed initial
radial velocity, $v_{z}=158$ km~s$^{-1}$, and sampling the
$(v_x,\,v_y)$ parameter space in steps of 5 km~s$^{-1}$ around the
value $(-80,\,132)$~km~s$^{-1}$.  Filled squares denote orbits that are found
to be acceptable fits to the stream data, whereas empty squares denote the
rest of the orbits in the grid.  Large symbols highlight the three cases chosen
for further analysis: the large squares represent the two extremes of the
acceptable set of orbits (i.e., orbits~`A' and `C' with inclinations $i=70^\circ$
and $115^\circ$ to the plane of the sky, respectively) and the large
hexagon represents a central case (orbit~`B', with $i\sim90^\circ$ to the plane
of the sky, i.e., an edge-on orbit).
}}
\end{center}
\end{figure}

All grid orbits are shown in Figure \ref{fig:grid} with empty square symbols.
Among these orbits we need to select those that fit the stream data.  Given that
stream data may not be an accurate representation for the progenitor data
(streams may deviate significantly from the progenitor's orbit), a
``best-fit'' method may not always be relevant.  Therefore, we choose to adopt
a simple accept-or-reject method by which we consider as acceptable only those
orbits that pass within a series of ``boxes'' centered on the stream data and
extending in both spatial and velocity dimensions.  A reasonable choice for the
size of these boxes would be of the order of the measurement uncertainties.
Figure \ref{fig:grid} shows with small filled symbols the orbits accepted
based on the criterion: $|\Delta x | = | \Delta y | = | \Delta z | = 16$~kpc
and $ | \Delta v_{z}| = 16$~km~s$^{-1}$.  Note that we consider boxes only
around fields~`1'--`8', since fields~`12'--`13' are generally difficult to
fit (see a similar discussion by \cite{iba04}).  From the set of acceptable
orbits we choose three to illustrate their common characteristics.
These are highlighted in Figure~\ref{fig:grid} with large symbols.  The large
squares represent the two~orbits with extreme inclinations out of the
``acceptable'' set ($i=70^\circ$ and $115^\circ$) and are denoted from now on
as orbits~`A' and `C', respectively.  The large hexagon represents
an intermediate case (edge-on to the plane of the sky---i.e., $i\sim90^\circ$),
and is denoted as orbit~`B'.  Note that orbit~`B' satisfies the most stringent
conditions: for example, by reducing the box sizes to $|\Delta x | =
| \Delta y | = | \Delta z | = 15$~kpc and $ | \Delta v_{z}| = 15$~km~s$^{-1}$,
the set of acceptable orbits consists of only a few centered around orbit~`B'.
Not surprisingly, the orbits found by us to be good fits to the stream data
are similar to the orbits presented by \citet{iba04} and \citet{lew05}.

Figure~\ref{fig:sim_orb1} shows orbits `A', `B' and `C' in more detail: The
top and middle panels show the $(x,\,y)$ and $(x,\,z)$ projections of the
stream data and of the orbits; the bottom panel shows the radial velocity
$v_z$ along the orbits compared with the observed values in fields~`1', `2',
`a3', `6', and `8', as well as the data on M31 satellite galaxies.
Figure~\ref{fig:sim_orb2} illustrates the three-dimensional positions of the
stream fields and of orbit `B' in the
$(x_{\rm M31},\,y_{\rm M31},\,z_{\rm M31})$ system of coordinates.

Note, we give more weight in our fit to radial velocities than to
line-of-sight distances because the latter are more susceptible to systematic
errors---e.g.,~contamination of the tip of the red giant branch region by
intermediate-age asymptotic giant branch stars, metallicity effects, etc..
The uncertainty in {\it relative\/} distance between M31 and the stream
fields may be larger than the 20~kpc distance error quoted by \citet{mcc03}
because of differences between the stellar populations of the stream and the
central region of the galaxy.  Finally, field~`8' has the weakest contrast of
the stream against the main body of M31 and this may be problematical for both
distance and radial velocity measurements \citep[e.g.,~see Fig.~1
of][]{iba04}.

\subsection{Implications}\label{sec:impl}

Several orbits provide good fits to the data, a reflection of the limited
nature of the observational constraints.  However, we found orbits with
a small range of inclinations to the plane of the sky ($i=70^\circ$--$115^\circ$)
to fit the giant southern stream data well.  All orbits that fit the data
share the common characteristics of a fairly high eccentricity and approximately
the same apocenter.  These are a consequence of the imposed constraints on the
orbit: The measurement of zero velocity (relative to M31) at the southern end
of the stream (field~`1'), coupled with the large range in line-of-sight distances
which tells us that the orbit is inclined to the line of sight, imply that
field~`1' must be near apocenter.  This working hypothesis is capable of being further
verified or falsified by observations: If it holds true, then the stream, as seen
in projection, should turn around on itself at or slightly beyond field~`1'.

\begin{figure}
\begin{center}
\epsfxsize=16cm\epsfysize=16cm\epsfbox{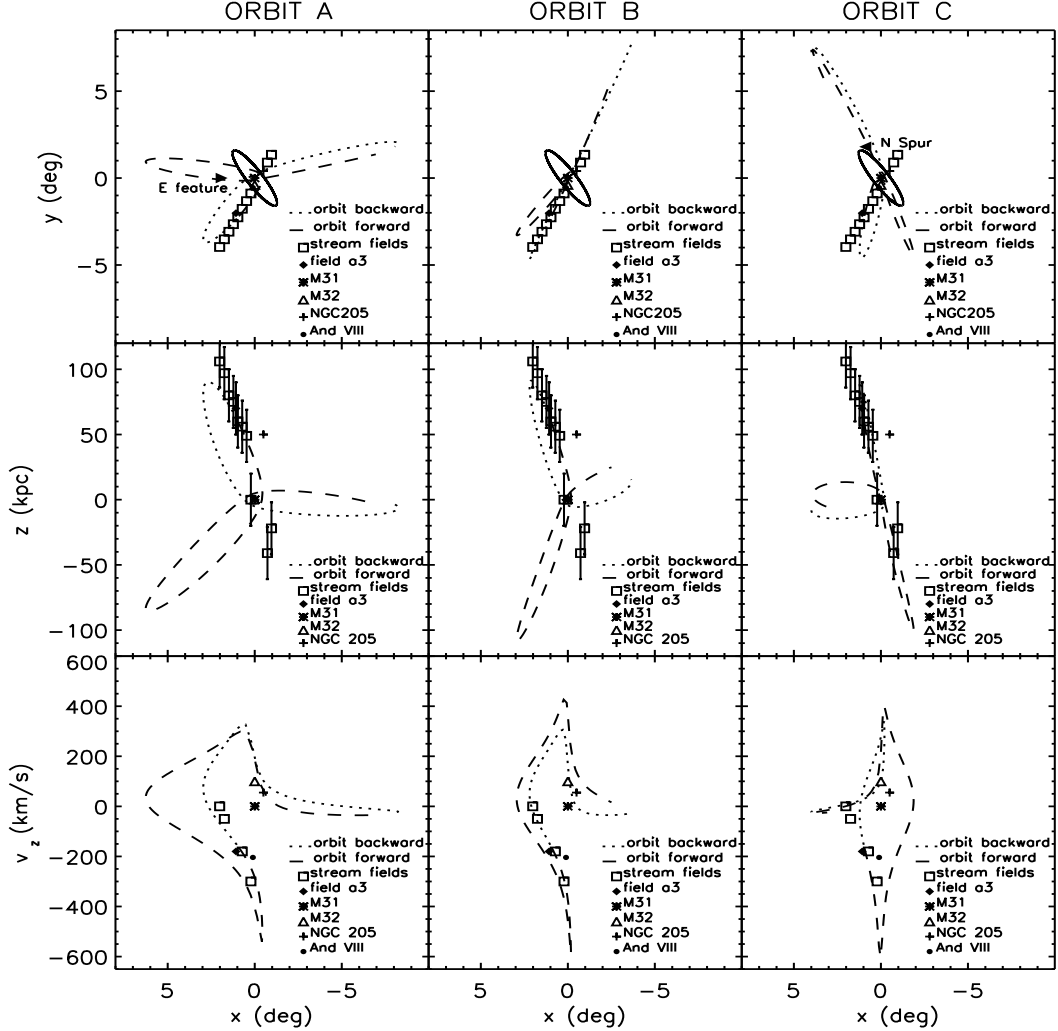}
\caption{\label{fig:sim_orb1}{
({\it Top panels\/})~The $(x,\,y)$-projection of the giant southern stream
fields, data on M31 satellite galaxies, and integrated orbits.  The three
columns correspond to the three selected orbits~`A', `B', and `C'.  The ellipse
delineates the limit of the visible disk of M31 (semi-major axis of $2^\circ$
or 27~kpc).~~
({\it Middle panels\/})~Same as above, but showing the $(x,\,z)$-projection.~~
({\it Bottom panels\/})~The radial velocity $v_z$ along the orbits plotted 
against the $x$-component of the distance.  The radial velocity measurements in
fields~`1', `2', `6', and `8' along the stream are approximate values
inferred from \citet{iba04}.
}}
\end{center}
\end{figure}

\begin{figure}
\epsfxsize=14cm\epsfysize=14cm\epsfbox{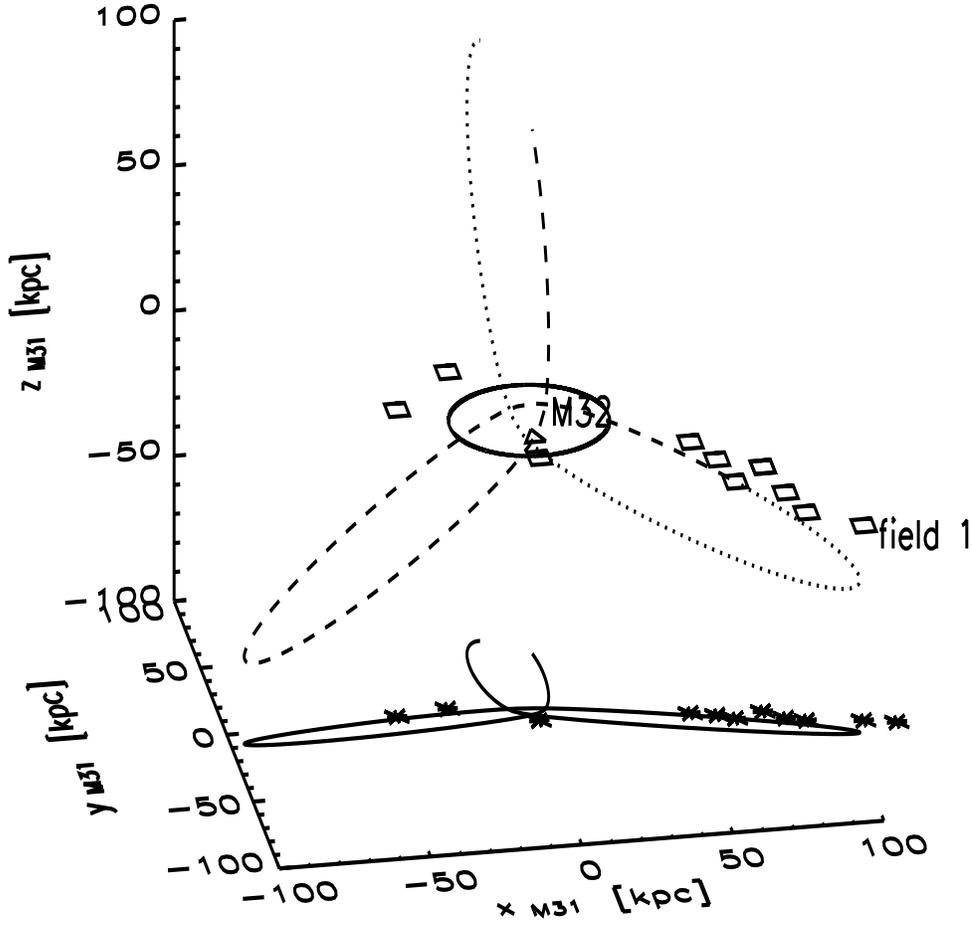}
\caption{\label{fig:sim_orb2}{
The three-dimensional position of the giant southern stream and orbit~`B'
(see \S\,2)
in the system of coordinates $(x_{\rm M31},\,y_{\rm M31},\,z_{\rm M31})$.
The dotted line denotes the orbit integrated backwards in time and the dashed
line denotes the orbit integrated forward in time.  The positions of M32 and
field~`1' are indicated.  The solid line and star symbols
illustrate the projection of the orbit and the stream field positions
onto the plane of the M31 disk, respectively.  The ellipse represents the
visible disk of M31, a circle of radius 27~kpc in the $(x_{\rm M31},\,y_{\rm
M31})$ plane.
}}
\end{figure}

Given the spatial and velocity constraints imposed by the data, we also
conclude from inspection of Figure \ref{fig:sim_orb1} that:

\begin{itemize}
\item[(1)]{It is generally difficult to fit the northern part of the stream
(i.e.,~fields~`12' and `13').  In order to fit these fields one needs a large
initial velocity; however, this would increase the eccentricity and apocenter
radius of the orbit rendering it inconsistent with the zero line-of-sight
velocity measured in field~`1'.}

\item[(2)]{We obtain a sub-set of orbits that pass close to the
northern spur feature (eg. orbit~`C').  Unfortunately, no velocity measurements
are avalaible at the moment in the northern spur in order to confirm or rule out
this association. (Note that before full spatial data and any
radial velocity were available, \citet{mer03} proposed an orbit that appears
to match both the stream and the northern spur.  This orbit can now be ruled
out because it has a turning point around field~`6' in the stream,
a location which is much closer to M31 than the current detections in field~`1'.
Also, a turning point in field~`6' implies a close to zero line-of-sight velocity
relative to M31 at that location, which is inconsistent with the current velocity
measurements in adjacent fields.)}

\end{itemize}

Among all orbits that fit the data, the orbit~`A' coincides, at least in
projection, with the eastern high-metallicity feature.  This feature, located at
$\xi\sim2^\circ$, $\eta\sim0^\circ$ in Fig.~5 of \citet{fer02}, can be either roughly
along the post-pericenter part of the orbit or just before the next pericenter of the
orbit.  The possible association with either of these two parts of the orbit can be
probed only by future velocity measurements in this region.  If the eastern
high-metallicity feature and the giant southern stream are indeed associated, we can
predict line-of-sight distances and radial velocities along this feature based on the
orbit determined in \S\,\ref{sec:part_orb}.  As can be inferred from
Figure~\ref{fig:sim_orb1}, this feature is expected to lie at a distance of
$\sim100$~kpc from M31 and in front of M31 as seen from our location.

\section{Constraints on the Progenitor Mass and Debris Age}\label{sec:width}

\subsection{Stream Width, Length, and Luminosity}\label{sec:width_stream}

Simple intuition tells us to expect debris from more massive satellites to
produce wider debris streams that spread more rapidly along the orbit with
time.  \citet*{joh01} present simple analytic scalings for the width and
length of debris streams, assuming the progenitor is a hot stellar system.
In this section, we use these ideas to discuss possible limits on the
characteristics of the progenitor.

\begin{figure}
\epsfbox{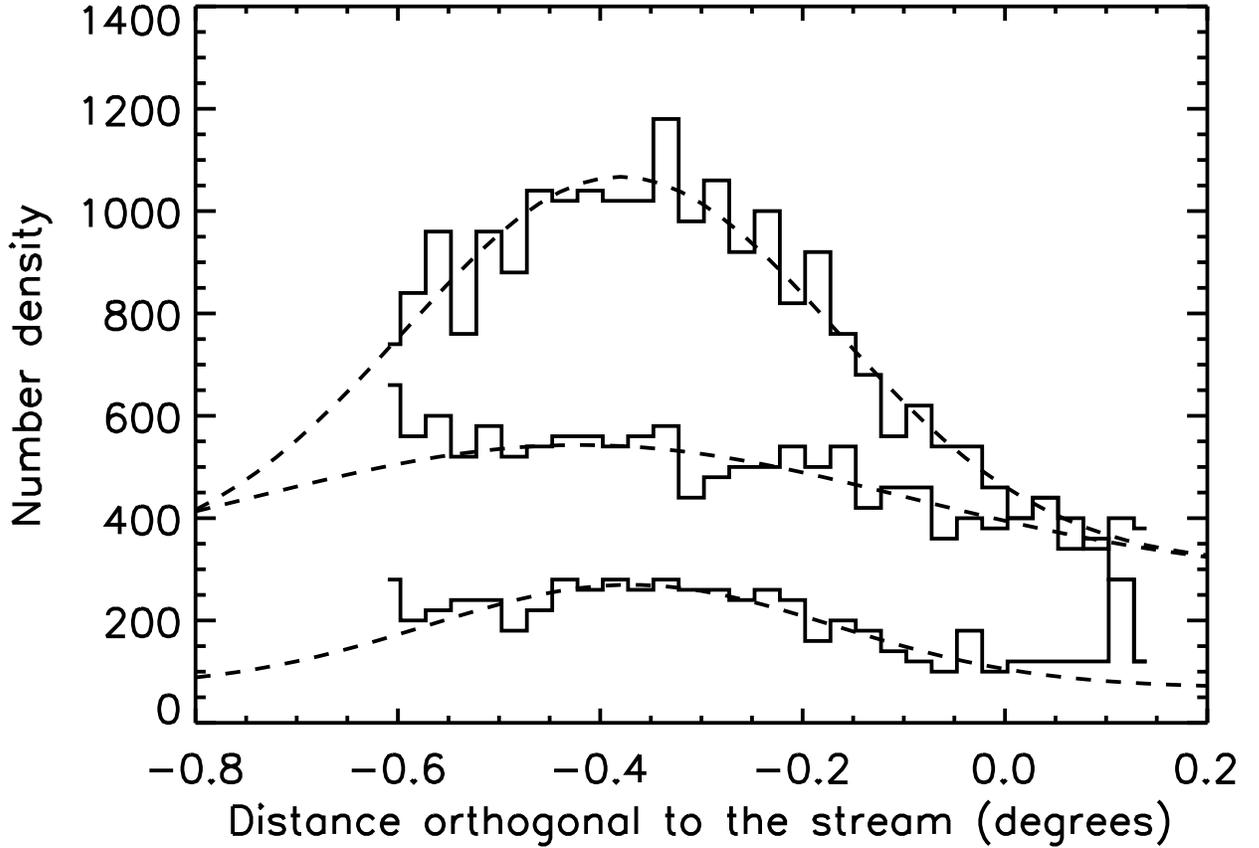}
\caption{\label{fig:sim_gauss}{
The distribution of stars at various locations along the stream, as measured
by \citet{mcc03}.  The top histogram corresponds to fields~`6'--`7', the
middle one to fields~`4'--`5', and the bottom one to fields~`1'--`3'.  The
dashed lines represent the best Gaussian fits to the data.  The width of the
stream (which is proportional to the standard deviation $\sigma$ of the
Gaussian) does not vary significantly along the southern fields.
}}
\end{figure}

Based on their star-count data, \citet{mcc03} suggest that the giant southern
stream is fanning out towards the southern part of the stream.  Numerical
simulations show that the stream is indeed expected to fan out towards the
apocenter of the orbit (in this case, towards field~`1').  To test this
expectation, we have estimated the width of the stream by fitting Gaussian
functions at various points along the stream to the density count data of
\citeauthor{mcc03}.  These fits are presented in Figure~\ref{fig:sim_gauss}.
Our results do not show a significant outward increase in the standard
deviation $\sigma$ of the Gaussians fitted to the stream, although the counts
are so low in the outer fields that we feel this cannot yet be ruled out.

Given the uncertainties in the other fields, we estimate the width of the
stream containing 80\% of the luminosity \citep[following][]{joh01} from the
combined fields~`6'--`7' to be $w=2.5\sigma\simeq0.5^\circ$, or, assuming an
average distance $d=833$~kpc, $w\simeq7.5$~kpc; the combined average distance
of these two~fields from the center of the galaxy is $R\approx58$~kpc.  These
parameters can be then used to put some limits on the mass of the progenitor
satellite at the time of disruption \citep[see a similar analysis
in][]{joh01}.  If the progenitor were a hot stellar system, the measured
fraction, $s\equiv{w/R}\simeq0.13$, is related to the mass $m$ of the
satellite through the relation:
$s=\Bigl[\frac{Gm}{(v_{\rm circ})^2R_{\rm peri}}\Bigr]^{1/3}$.  Given that
$v_{\rm circ}=260$~km~s$^{-1}$ and $R_{\rm peri}$ is in the range
3$\,$--$\,$4.5~kpc for our accepted orbits, one can estimate the mass of the satellite
prior to the disruption from which the debris came: $m\sim1.0\,$--$\,1.6\times10^8
\,M_\odot$.  From the luminosity of the stream so far detected, 
$L=3\times10^7L_\odot$ \citep{iba01a}, this gives mass-to-light
ratios of~3.5--5.2, which are marginally consistent with those
inferred from observations of nearby {\it dwarf elliptical\/} galaxies.  The low
$M/L$ ratio may be indicative of the nature of the progenitor.
However, we caution that this ratio is rather uncertain: The progenitor may
not be completely disrupted or the observed luminosity may not be representative
for the entire stream, as we have implicitly assumed.  Also, for a system of similar
mass with some rotation, the associated streams would be thinner, in which
case these are lower bounds on the mass and mass-to-light ratios of the
progenitor. These derived constraints on the progenitor's properties
   are similar to those found by \citet{iba04}.

In addition, the time $t$ taken for the observed debris to spread along the
orbit can be estimated from its observed angular extent $\Psi$ around the
parent galaxy.  \citet{joh01} write down an expression, most valid for mildly
eccentric orbits, that uses $\Psi/2\pi$ directly as an estimate of the
fraction of the orbit covered with debris.  Since our orbit is highly
eccentric we adapt their equation~(5), replacing $\Psi/2\pi$ with
$\Delta{t}/T_\Psi$, where $\Delta{t}$ is the time taken to travel $\Psi$
along the orbit and $T_\Psi$ is the azimuthal period:
\begin{equation}
{\Delta{t}\over{T}_\Psi}\simeq4s\frac{t}{T_\Psi}~~.
\label{eqn:t_psi}
\end{equation}
From our test particle orbits we find $\Delta{t}=0.18~-0.21$~Gyr, and hence 
the time since disruption for the giant southern stream is $t=0.35 - 0.4$~Gyr. 
In conclusion, our results suggest that the giant southern stream in M31 is 
very young (less than an orbit old; for this orbit, the azimuthal periods for 
the accepted orbits are $T_\Psi \simeq 1.65$~Gyr and the radial orbital 
periods are $T_R \sim 1.45$~Gyr). This result is also supported by the visual 
appearance of the stream: The stream is young enough to be plainly visible 
as an overdensity in star counts \citep{joh98}.

\subsection{The Coldness of the Stream}\label{sec:cold}

In the companion paper (Paper~I), the intrinsic line-of-sight velocity
dispersion of the giant southern stream in field~`a3' was constrained to be
$\lesssim23$~km~s$^{-1}$.  What can be inferred from the coldness of the
stellar stream?

\citet{hel99} demonstrate that the dispersion in debris should decrease over
time.  However, since the giant southern stream is very young, we do not
expect dynamical cooling to be significant yet.  Also, neither has it had
time yet to be significantly heated by tidal interactions with dark matter
substructure in the halo \citep*{jsh02}.  Rather, the velocity dispersion is
expected to vary most significantly in an oscillatory manner as a function of
radial orbital phase \citep[a result also obtained by][]{hel99}.

\begin{figure}
\epsfxsize=12cm\epsfysize=16.5cm\epsfbox{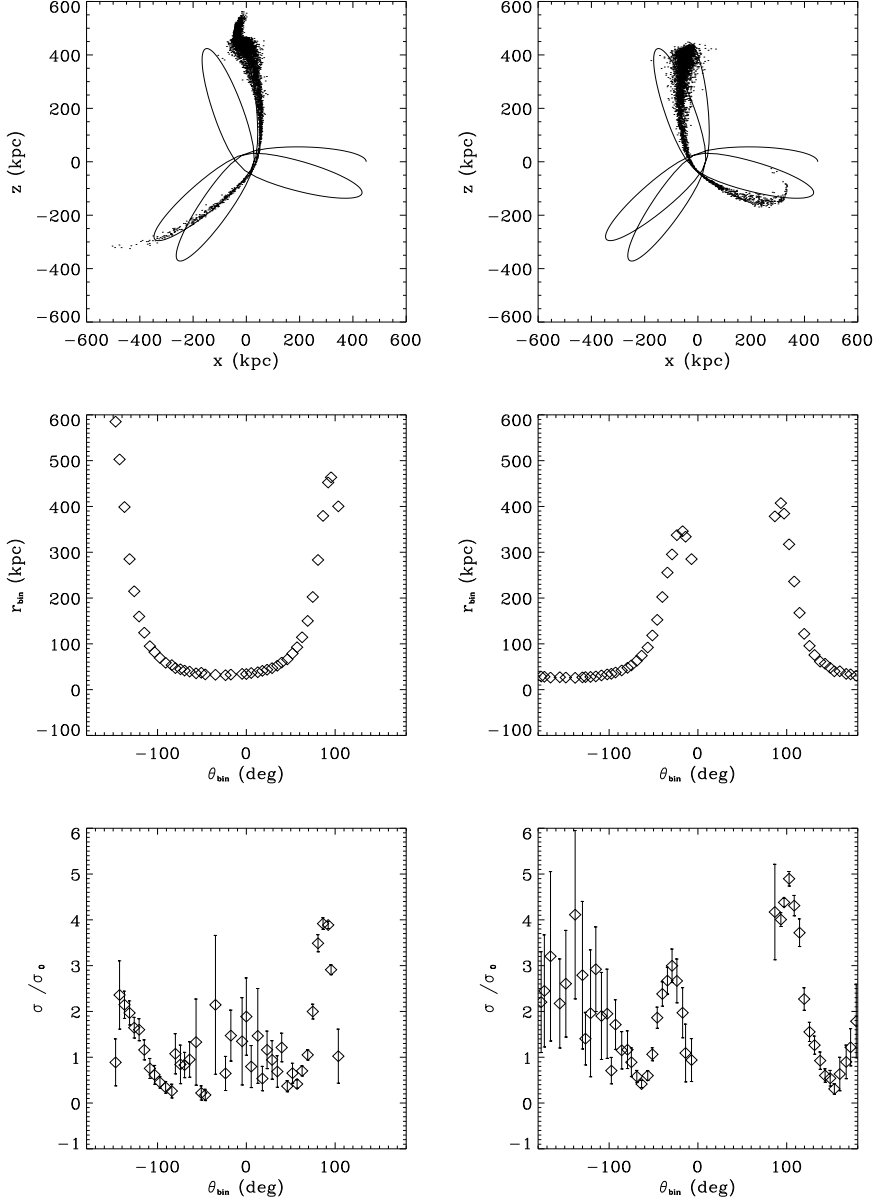}
\caption{\label{fig:sim_sigma}{
Results of an $N$-body simulation of a stellar stream moving in a highly
eccentric orbit, similar to that of the orbits determined in \S\,\ref{sec:orb}.
The left and right panels correspond to the trailing and
leading portions of the stream, respectively.~~
({\it Top\/})~Positions of test particles in the orbital plane at
apocenter.  The solid line shows the orbit of the satellite prior to that time.~~
({\it Middle\/})~Distance from the center of the parent galaxy plotted
against the azimuthal angle $\theta$ along the giant southern stream.~~
({\it Bottom\/})~Radial velocity dispersion with respect to the center of the
parent galaxy in units of the central dispersion of the satellite,
$\sigma_0$, plotted versus the azimuthal angle $\theta$ along the stream.
The model suggests that, along a stellar stream moving on a highly eccentric
orbit, the velocity dispersion may vary drastically relative to that of the
progenitor galaxy.
}}
\end{figure}

Based on the orbit determined in \S\,\ref{sec:part_orb}, we can make some
general comments about the phase-space dependence of the giant southern
stream.  As discussed before, several arguments suggest that the orbit of the
stream is highly eccentric.  The orbits presented in Figure~\ref{fig:sim_orb1}
have apocenter/pericenter ratios ranging $\frac{R_{\rm apo}}{R_{\rm
peri}}\sim 25\,$--$\,$30.  In Figure~\ref{fig:sim_sigma} we show the result of an $N$-body
numerical simulation of a stellar stream moving in an orbit of high
eccentricity ($\frac{R_{\rm apo}}{R_{\rm peri}} \sim15$), as an illustration
for the trends expected to occur along eccentric orbits;
this is ``Model~4'' in the numerical simulations of \citet*{jcg02}.  The top
two~panels of Figure~\ref{fig:sim_sigma} show the position in the orbital
plane of the stellar debris in both the trailing (left panel) and leading
(right panel) portions of the stream.  The middle panels show the distance
from the center of the parent galaxy versus azimuthal angle $\theta$ along
the stream.  The bottom panels show the radial velocity dispersion (with
respect to the parent galaxy) in units of the central dispersion of
satellite, $\sigma_0$, versus azimuthal angle $\theta$ along the stream.
Note, because the orbital plane of the stream is almost edge on to our line
of sight we expect the observed radial velocity dispersion (i.e.,~along the
line of sight) for fields~`2'--`8' in the giant southern stream to exhibit similar
effects as the radial velocity dispersion with respect to parent galaxy in the
simulations.  The parameters are calculated as averages over all particles in
uniform bins in $\theta$.

From Figure~\ref{fig:sim_sigma} one can infer some general trends about the
phase-space evolution of the debris: Spikes in $\sigma/\sigma_0$, as large as
a factor of~4--5, occur at the turning points (pericenter and apocenter),
as predicted by \citet{hel99}. Also, the stream can become very cold in between
the turning points with the velocity dispersion of the stream reaching values
well below the central dispersion of the satellite, as small as
$\sigma/\sigma_0\sim0.5$. These effects are most pronounced for and
appear to be generic features of more eccentric orbits (simulations with less 
eccentric orbits are not shown here). Therefore these results could be 
even higher for the orbit of the progenitor, given that the orbits found 
in \S \ref{sec:part_orb} are even more eccentric than those of our numerical 
simulations. 

Our field~`a3' lies between apocenter and pericenter in orbital phase along
our fitted orbits, and hence we expect the velocity dispersion of the
progenitor to be at least as large as the intrinsic value estimated for the 
stream, and possibly much greater than this. Adopting the nominal best-fit 
value of 15~km~s$^{-1}$ from the possible range of stream velocity dispersions
(0\,--\,23~km~s$^{-1}$), this implies that the progenitor satellite has a
mass of $\gtrsim10^8M_\odot$.  This lower bound, admittedly a rough one given
the caveats discussed above, is consistent with that set by the width of the
debris on the mass of the satellite (see \S\,\ref{sec:width}).

Future velocity measurements can be used to confirm or rule out our prediction
of significant variations of the line of sight velocity dispersion along the giant
southern stream.  \citet{iba04} do have some velocity measurements in these
fields, but not in sufficient numbers to look for this effect; combining data
from fields~`1', `2', `6', and `8', they find a concentration of stars with a
velocity dispersion of 11~km~s$^{-1}$, but with a skewed tail of velocities
relative to the center of the stream with a spread much greater than this
(possibly related to M31's smooth halo population; see Paper~I).

\section{Discussion: The Possible Progenitor}\label{sec:disc}

From the two~independent mass estimates described above
(\S\,\ref{sec:width_stream} and \S\,\ref{sec:cold}) we conclude that our
progenitor satellite has a mass $m>10^8M_\odot$ but this is only a rough
lower bound. This result suggests that the progenitor is a massive dwarf
galaxy.  In particular, we note that although the observational data give an
upper limit to the velocity dispersion of the stream, the theoretical models
show that the stream's velocity dispersion provides only a lower limit to the 
velocity dispersion of the progenitor.  Therefore, the satellite can have a 
velocity dispersion much larger than 23~km~s$^{-1}$.  Consistent with this 
result are the new measurements of the mean metallicity of red giant stars in 
field~`a3' in the giant southern stream (Paper~I): 
$\rm\langle[Fe/H]\rangle=-0.5$~dex.  Assuming that this value is
representative for the satellite as well, and using the empirical
metallicity-luminosity relation obtained in the Local Group
\citep{mat98,dek03}, this would imply an absolute magnitude $M_B=-17$
($L_B\sim10^9L_\odot$) and mass $m\approx5\times10^9M_\odot$ for the
progenitor satellite.  While the metallicity-based estimates of the
progenitor mass and luminosity are much greater than the lower bounds
obtained from the stream width, luminosity, and velocity dispersion, it
should be recognized that those lower bounds are very approximate for the
reasons discussed above.

The large discrepancy between the direct estimate of the stream's luminosity
\citep{iba01a,mor03} and the progenitor luminosity inferred from our 
metallicity measurement may have an important implication.  The former
estimate corresponds only to the detected part of the stream---the luminosity
of the {\it entire\/} stream may be much higher.  This would suggest that a
large portion of the stream or even its progenitor are currently invisible
(or unidentified), either because a part of the stream has already faded into
the background and/or the surviving portions of the satellite and stream are
lost against the disk of M31.  The recent identification of a high-luminosity
feature ($L\sim10^8L_\odot$) along the stream \citep {mor03} should caution
us that other stream features may still remain undetected.

Several satellites that fit these mass and luminosity descriptions are
aligned in projection along or in the close vicinity to the giant southern
stream.  The velocity information for the debris, which has become available
only recently, is useful for ruling out some of these possible associations.

Based on our orbit integrations we can conclude that M32 is unlikely to be
associated with the stream. Although M32 has a projected position almost
coincident with the stream, its radial velocity has an opposite sign to those
of the velocities of the fields in the giant southern stream.  We can
therefore exclude any possibility that the stream results from a {\it current}
passage of M32.  This can be clearly seen in the radial velocity plot in the
bottom panel of Figure~\ref{fig:sim_orb1}: The radial velocity starts 
from field `a3' with a value of $-158$~km~s$^{-1}$ and decreases during the 
first passage; it becomes positive only in a subsequent passage. Several 
authors mention the possibility that the observed stream could be a remnant 
from a previous passage of M32 \citep[e.g.,][]{iba04}.  We believe that this 
is also inconsistent with the observations---if this scenario were true, a 
stream  from the current passage should also be visible, and this is not 
supported by observations.  NGC~205 is also an unlikely progenitor based on 
similar velocity arguments, as well as on the fact that its line-of-sight 
distance does not match our fitted orbits \citep[see also the result of][]{iba04}.

The satellite responsible for the stream should be currently located along
the stream. The satellite can either be one of the surviving satellites
around M31 or it could be totally destroyed. Given the relatively young age
of the stream, the latter case implies that the satellite in question was
destroyed only a short time ago.  The recently discovered And~VIII
\citep{mor03} is an attractive possibility as a progenitor: Both its location
and radial velocity ($\xi\sim0^\circ$, $\eta\sim-0.5^\circ$; $v_{\rm
rad}=-204$~km~s$^{-1}$ with respect to M31) are consistent with those of the
giant southern stream and of our determined orbit (see~Fig.~\ref{fig:sim_orb1}).
Future observations need to confirm if And~VIII is a satellite galaxy or,
as has been recently suggested \citep{iba04}, simply a part of the stellar stream.

\section{Summary}\label{sec:summary}

From a comparison of test particle orbits with observational data we conclude
that the progenitor of the M31 giant southern stream was (or is) on a highly
eccentric, close to edge-on (to the plane of the sky) orbit with apo- to
peri-center ratio of order~25--30, and an apocenter at or only slightly beyond
the edge of the current data.  Given these accepted orbits we estimate the mass
of the progenitor to be $>10^8M_\odot$ from the width of the debris, and the time
since disruption to be $0.25$~Gyr (less than one~orbit). Moreover, $N$-body
simulations suggest that our line-of-sight velocity dispersion limit of 23~km~s$^{-1}$
for the stream in field~`a3' is only a lower bound on the dispersion of the progenitor.
In conclusion, our analyses lead us to expect that:~~ (i)~The stream should
turn around slightly beyond the edge of field~`1';~~ (ii)~The stream should
widen around this turning point and the line -of-sight velocity dispersion should
exhibit significant variations along the stream;~~ (iii)~There are possible
associations between the post-pericenter part of the stream and the northern spur or
the eastern high-metallicity feature. In the latter case, this eash high metallicity
feature should lie well in front of the disk; and~~ (iv)~the kinematic data on And~VIII
are consistent with this feature being associated with the orbit of the giant
southern stream.

\acknowledgments

K.V.J.\ and A.F.'s contributions were supported through NASA grant NAG5-9064
and NSF CAREER award AST-0133617.  P.G.\ acknowledges support from NSF grant
AST-0307966 and a Special Research Grant from UCSC.  S.R.M.\ acknowledges
funding by NSF grants AST-0307842 and AST-0307851, NASA/JPL contract 1228235,
the David and Lucile Packard Foundation, and The F.~H.~Levinson Fund of the
Peninsula Community Foundation.  R.M.R.\ acknowledges support from NSF grant
AST-0307931.

\clearpage

\begin{thebibliography}{}

\bibitem[Bekki et~al.(2001)]{bekki01} Bekki, K., Couch, W. J., Drinkwater, M. J., \& Gregg, M. D. 2001, \apj, 557, L39
\bibitem[Dekel \& Woo(2003)]{dek03} Dekel, A.~\& Woo, J.\ 2003, \mnras, 344,
    1131
\bibitem[Ferguson et~al.(2002)]{fer02} Ferguson, A.~M.~N., Irwin, M.~J.,
    Ibata, R.~A., Lewis, G.~F., \& Tanvir, N.~R.\ 2002, \aj, 124, 1452
\bibitem[G\'{o}mez-Flechoso, Fux, \& Martinet(1999)]{gom99} G\'{o}mez-Flechoso, M. A., Fux, R.\& Martinet, L. 1999, \aap, 347, 77
\bibitem[Guhathakurta et~al.(2005)]{guh05} Guhathakurta, P., Rich, R.~M.,
    Reitzel, D.~B., Cooper, M.~C., Gilbert, K., Majewski, S.~R., Ostheimer,
    J.~C., Geha, M.~C., Johnston, K.~V., \& Patterson, R.~J.\ 2005, \aj,
    submitted (Paper~I; astro-ph/0406145)
\bibitem[Helmi \& White(1999)]{hel99} Helmi, A., \& White, S.~D.~M.\ 1999,
    \mnras, 307, 495
\bibitem[Helmi \& de~Zeeuw(2000)]{hel00} Helmi, A., \& de~Zeeuw, T.\ 2000,
    \mnras, 319, 657
\bibitem[Hernquist(1990)]{hernq90} Hernquist, L. 1990, \apj, 356, 359
\bibitem[Hurley-Keller et~al.(2004)]{hur04} Hurley-Keller, D., Morrison,
    H.~L., Harding, P., \& Jacoby, G.~H.\ 2004, \apj, 616, 804
\bibitem[Ibata et~al.(2004)]{iba04} Ibata, R., Chapman, S., Ferguson,
    A.~M.~N., Irwin, M., Lewis, G., \& McConnachie, A.\ 2004, \mnras, 351, 117
\bibitem[Ibata et~al.(1994)Ibata, Gilmore, \& Irwin]{iba94} Ibata, R.~A.,
    Gilmore, G., \& Irwin, M.~J.\ 1994, \nat, 370, 194
\bibitem[Ibata et~al.(2001a)]{iba01a} Ibata, R., Irwin, M.~J., Ferguson,
    A.~M.~N., Lewis, G., \& Tanvir, N.\ 2001, Nature, 412, 49
\bibitem[Ibata \& Lewis(1998)]{iba98} Ibata, R.~A., \& Lewis, G.~F.\ 1998,
    \apj, 500, 575
\bibitem[Ibata et~al.(2001b)]{iba01b} Ibata, R.~A., Lewis, G.~F., Irwin,
    M.~J., Totten, E., \& Quinn, T.\ 2001, \apj, 551, 294
\bibitem[Johnston(1998)]{joh98} Johnston, K.~V.\ 1998, \apj, 495, 297
\bibitem[Johnston et~al.(2002a)Johnston, Choi, \& Guhathakurta]{jcg02}
    Johnston, K.~V., Choi, P.~I., \& Guhathakurta, P.\ 2002a, \aj, 124, 127
\bibitem[Johnston et~al.(2005)Johnston, Law, \& Majewski]{joh05} Johnston,
    K.~V., Law, D.~R., \& Majewski, S.~R. 2005, \apj, in press (astro-ph/0407565)
\bibitem[Johnston et~al.(1999)]{joh99} Johnston, K.~V., Majewski, S.~R.,
    Siegel, M.~H., Reid, I.~N., \& Kunkel, W.~E.\ 1999, \aj, 118, 1719
\bibitem[Johnston et~al.(2001)Johnston, Sackett, \& Bullock]{joh01} Johnston,
    K.~V., Sackett, P.~D., \& Bullock, J.~S.\ 2001, \apj, 557, 137
\bibitem[Johnston et~al.(2002b)Johnston, Spergel, \& Haydn]{jsh02} Johnston,
    K.~V., Spergel, D.~N., \& Haydn, C.\ 2002b, \apj, 570, 656
\bibitem[Johnston et~al.(1995)Johnston, Spergel, \& Hernquist]{joh95}
    Johnston, K.~V., Spergel, D.~N., \& Hernquist, L.\ 1995, \apj, 451, 598
\bibitem[Kent(1989)]{ken89} Kent, S.~M.\ 1989, \pasp, 101, 489
\bibitem[Law et~al.(2005)Law, Johnston, \& Majewski]{law05} Law, D.~R.,
    Johnston, K.~V., \& Majewski, S.~R.\ 2005, \apj, in press (astro-ph/0407566)
\bibitem[Lewis et~al.(2005)]{lew05} Lewis, G.~F., Ibata, R.~A., Chapman,
    S.~C., Ferguson, A.~M.~N., McConnachie, A.~W., Irwin, M.~J., \& Tanvir,
    N.\ 2005, in 5th Galacto Chemodynamics conference, PASA, in press
    (astro-ph/0401092)
\bibitem[Majewski et~al.(2003)]{maj03} Majewski, S.~R., Skrutskie, M.~F.,
    Weinberg, M.~D., \& Ostheimer, J.~C.\ 2003, \apj, 599, 1082
\bibitem[Mateo(1998)]{mat98} Mateo, M.~L.\ 1998, \araa, 36, 435
\bibitem[McConnachie et~al.(2003)]{mcc03} McConnachie, A.~W., Irwin, M.~J.,
    Ibata, R.~A., Ferguson, A.~M.~N., Lewis, G.~F., \& Tanvir, N.\ 2003,
    \mnras, 343, 1335
\bibitem[Merrett et~al.(2003)]{mer03} Merrett, H.~R., et~al.\ 2003, \mnras,
    346, L62
\bibitem[Merrett et~al.(2005)]{mer05} Merrett, H.~R., et~al.\ 2005, in Proc.\
    of ESO Workshop: Planetary Nebulae beyond the Milky Way, ed. J.~R.~Walsh
    \& L.~Stanghellini (Springer-Verlag), in press (astro-ph/0407331)
\bibitem[Miyamoto \& Nagai(1975)]{miya75} Miyamoto, M. \& Nagai, R.\ 1975, \pasj, 27, 533
\bibitem[Morrison et~al.(2003)]{mor03} Morrison, H.~L., Harding, P.,
    Hurley-Keller, D., \& Jacoby, G.\ 2003, \apjl, 596, L183
\bibitem[Tremaine(1993)]{tre93} Tremaine, S.\ 1993, AIP Conf.\ Proc.~278:
    Back to the Galaxy, 599
\bibitem[Vel{\'{a}}zquez \& White(1995)]{vel95} Vel{\'{a}}zquez, H., \&
    White, S.~D.~M.\ 1995, \mnras, 275, 23
\end{thebibliography}
\end{document}